\documentclass[10pt,twocolumn]{article} 
\usepackage{simpleConference}
\usepackage{times}
\usepackage{graphicx}
\usepackage{amssymb}
\usepackage{url,hyperref}
\hypersetup{hypertex=true,
	colorlinks=true,
	linkcolor=blue,
	anchorcolor=blue,
	citecolor=blue,
	urlcolor=blue,
}
\usepackage{amsmath}
\usepackage{cite}
\usepackage[margin=10pt,font=small,labelfont=bf,
 labelsep=endash]{caption}
\begin{document}

\title{Ultrafast two-photon emission in a doped semiconductor thin film}

\author{Iroro Orife \\
\\
Technical Report \\
Seattle, Washington, USA \\
\today
\\
\\
iroro@alumni.cmu.edu  \\
}
\author{Futai Hu,$^{1}$ Liu Li,$^{1}$ Yuan Liu,$^{1}$ Yuan Meng,$^{1}$ Mali Gong,$^{1, 2, \ast}$ Yuanmu Yang$^{1, \dagger}$\\
$^1$State Key Laboratory of Precision Measurement Technology and Instruments,\\ Department of Precision Instrument, Tsinghua University, Beijing 100084, China\\
$^2$State Key Laboratory of Precision Measurement Technology and Instruments,\\ Department of Precision Instrument, Tsinghua University, Beijing 100084, China\\
\today
\\
${\dagger}$ymyang@tsinghua.edu.cn\\
${\ast}$gongml@mail.tsinghua.edu.cn\\
}%

\maketitle
\thispagestyle{empty}

\begin{abstract}
		As a high-order quantum transition, two-photon emission has an extremely low occurrence rate compared to one-photon emission, thus having been considered a ``forbidden'' process. Here, we propose a scheme that allows ultrafast two-photon emission, leveraging highly confined surface plasmon polariton modes in a degenerately-doped, light-emitting semiconductor thin film. The surface plasmon polariton modes are tailored to have simultaneous spectral and spatial overlap with the two-photon emission in the semiconductor. Using degenerately-doped InSb as the prototype material, we show that the two-photon emission can be accelerated by 10 orders of magnitude: from tens of milliseconds to picoseconds, surpassing the one-photon emission rate. Our result provides a semiconductor platform for ultrafast single and entangled photon generation, with a tunable emission wavelength in the mid-infrared.
\end{abstract}

\section{Introduction}
Two-photon emission (TPE) refers to the simultaneous emission of two photons during a quantum radiative transition\cite{Gauthier1992,He1994,Ning2004,Lissandrin2004,Hayat2007,Hayat2008,Hayat2009,Nevet2010-PRL,Nevet2010-NL,Lin2010,Ota2011,Poddubny2012,Rivera2017,Melzer2018}. Recent researches suggest TPE as a promising approach to generate entangled photon pairs in semiconductors, as it emits two photons with intrinsic energy conservation and time coincidence\cite{Hayat2011}. Compared to spontaneous parametric down conversion, a prevailing method to generate entangled photons, TPE does not have the restriction on phase-matching, and can potentially achieve a 3-order higher occurrence probability\cite{Kwiat1995,Hayat2008}. TPE can also occur within a wide temperature range, unlike semiconductor quantum dots that require low temperatures to efficiently generate entangled photons\cite{Huber2018,Chen2018,Liu2019}. In addition, while one-photon emission (OPE) during interband transition can only emit photons with energy above the material bandgap, TPE spectrum can be extremely broad starting from the near-zero frequency, which indicates that TPE has the potential to provide emission and gain with an ultra-broad bandwidth not restricted by the material bandgap. However, as a second-order quantum transition, the TPE rate is typically 5-10 orders of magnitude lower than the OPE rate, mainly due to the mismatch between the characteristic emitter size and the light-emitting wavelength\cite{Hayat2008,Hayat2009, Rivera2016,Rivera2017}. Consequently, the TPE rate needs to be dramatically increased, to meet the requirements for practical device applications.

Spontaneous emission rate can be increased by applying a specified dielectric environment, widely known as the Purcell effect\cite{Purcell1946}. The enhancement originates from an increased field confinement and density of states (DOS)\cite{Iwase2010}. The enhancement factor, i.e. Purcell factor can be given by $F = {\varGamma}/{\varGamma}^{0}$, where ${\varGamma}$ and ${\varGamma}^{0}$ are the modified and vacuum emission rate, respectively. Surface plasmon polaritons (SPPs), with broadband field confinement and high DOS \cite{Iwase2010}, have been applied to improve the emission rate of various light sources, such as light-emitting diodes and quantum dots\cite{Gontijo1999,Okamoto2004,Tanaka2010,Sauvan2013,Khurgin2014,Hoang2015,Caligiuri2018,Chen2018}. Since TPE and OPE have different spectral spans, Purcell effect can also be employed to selectively obtain an increased TPE rate\cite{Nevet2010-NL,Rivera2017}. Previous researchers have demonstrated  enhancements of the TPE rate in semiconductors by coupling emitters to a plasmonic bowtie nanoantenna array\cite{Nevet2010-NL}. However, the experimentally estimated TPE intensity is enhanced by a 3 orders of magnitude, yet still much lower than the OPE rate. This is likely due to that the optical field is only enhanced in the vicinity of the antenna tips and at the resonance frequency of nanoantennas. In addition, the spontaneous emission enhancement near semiconductor emitters is limited by the relatively low quality- ($Q$-) factor of the antenna resonance. More recently, Rivera \textit{et al.} theoretically proposed alternative approaches to enhance the TPE rate, with a calculated rate even surpassing the OPE rate, by placing an atomic emitter near a single-layer graphene that supports plasmons, or a polar dielectric film that supports surface phonon polaritons \cite{Rivera2016,Rivera2017}. However, a spatial separation between emitters and the field maximum still pose a limit to the overall TPE rate. An integrated light-emitting scheme that is straightforward for the experimental implementation is yet identified. We note that, strictly speaking, the term "two-photon emission" shall be replaced by "two-polariton emission" when considering emission into polariton modes. However, for simplicity, here we use these two terms interchangeably. 

\section{Principle}
Here, we propose a new scheme to achieve ultrafast TPE, by employing a degenerately-doped semiconductor thin film that simultaneously serves as the light-emitting medium and supports SPP modes. Doped semiconductors are known to support SPP modes in the mid- and far-infrared\cite{Wagner2014,Yang2017}. In this work, we design the modal
frequency of SPP modes to be around half of the OPE frequency of the semiconductor, thus allowing selective enhancement of the TPE rate in the direct bandgap semiconductor. A remarkable Purcell factor up to $1.2\times10^{6}$ is obtained in the prototype degenerately-doped InSb film.

\begin{figure}[!b]
	\begin{center}
	\includegraphics[width=7.5cm]{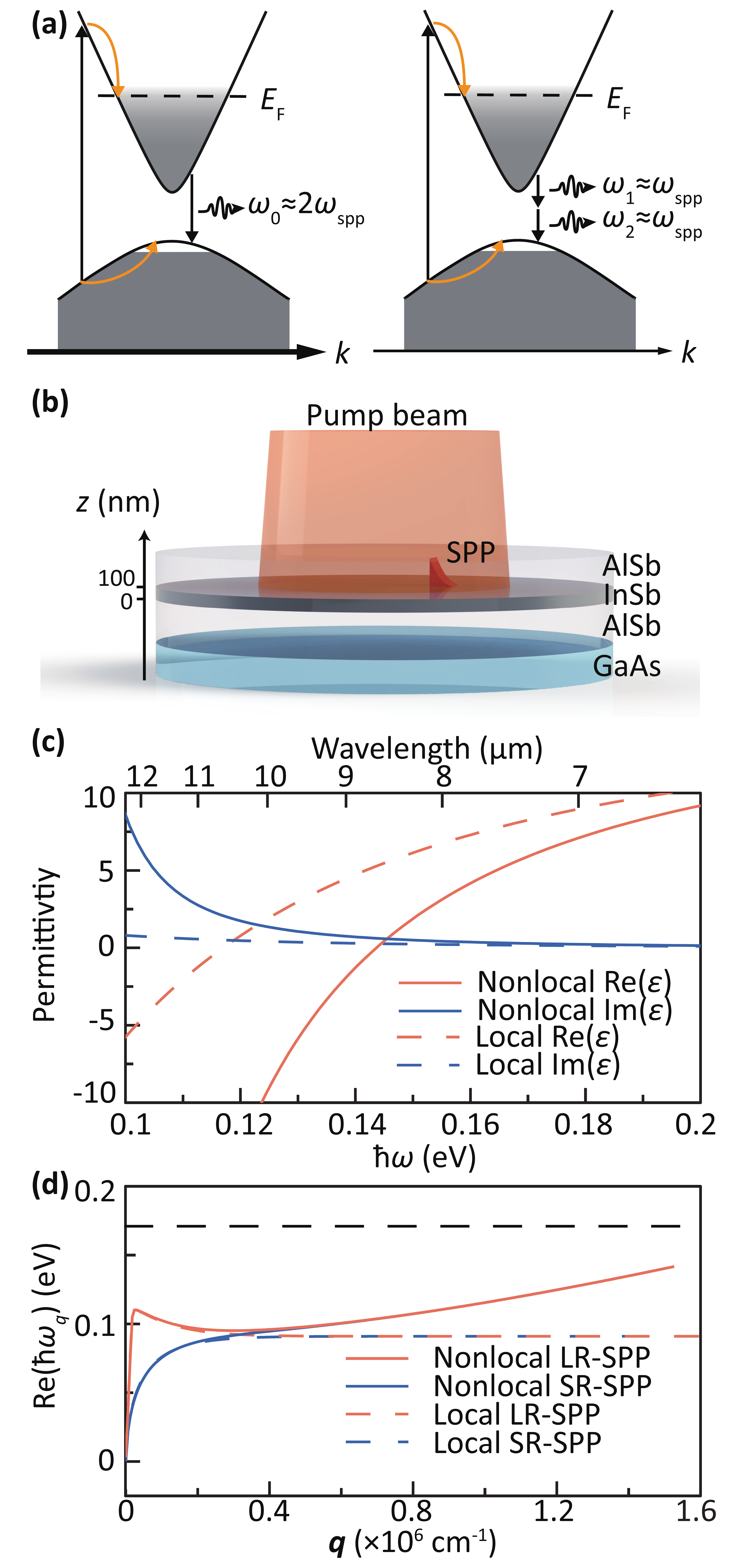}
	\end{center}
	\caption{\label{Fig. 1} (a) Schematics of OPE (left) and TPE (right) in a semiconductor with a degenerate doping level. Dashed line marks the Fermi level $E_{\rm F}$. Yellow arrows represent the thermalization of photoexcited carriers. (b) Schematic diagram of the prototype device. (c) Dispersion of the local and non-local permittivity of InSb with $N_{\rm e} = 8\times10^{18}$ cm$^{-3}$. (d) Local and nonlocal dispersion of LR-SPP and SR-SPP modes in the InSb film. The black dashed line marks the direct bandgap energy of InSb.}
\end{figure}

We theoretically demonstrate that the TPE can be enhanced by 10 orders of magnitude: accelerated from tens of millisecond to picoseconds, even faster than the OPE rate near the surface of the InSb layer. In addition, the TPE spectral	peak can be flexibly tuned by varying the doping density of the InSb film.

Fig.\ref{Fig. 1}(a) and \ref{Fig. 1}(b) show the schematics of OPE and TPE in the n-doped semiconductor, where the OPE frequency is designed to be about twice the SPP modal frequency of the same semiconductor thin film. The OPE and TPE rate, tailored by the Purcell effect, can be formulated as\cite{Nevet2010-NL,Rivera2017}, 
\begin{equation}
R_{\rm OPE}(\omega_{0}) = F(\omega_{0})R_{\rm OPE}^{0}(\omega_{0})%
\label{Eq1}
\end{equation}
\begin{equation}
R_{\rm TPE}(\omega_{1}, \omega_{2}) = F(\omega_{1})F(\omega_{2})R_{\rm TPE}^{0}(\omega_{1}, \omega_{2})
\label{Eq2}
\end{equation}
where $R_{\rm OPE}^{0}(\omega_{0})$ and $R_{\rm TPE}^{0}(\omega_{1}, \omega_{2})$ represent the OPE and TPE rate in vacuum, respectively\cite{SI}.

Our proposed structure comprises of a 100-nm-thick InSb film sandwiched by two AlSb layers, as shown in Fig.\ref{Fig. 1}(c). The bottom AlSb can serve as a buffer layer between GaAs and InSb for an optimal electron mobility of InSb\cite{Kang2018}. It can also serve as a wide-bandgap barrier to prevent carrier injection to the GaAs substrate. AlSb layers can be considered semi-infinite in the calculation due to the high confinement of SPP modes. The layered structure can support SPP modes, and the dispersion relation of these modes can be described as\cite{Wendler1986},
\begin{equation}
{1+\frac{({\varepsilon_1}{\alpha_2} - {\varepsilon _2}{\alpha _1})({\varepsilon _2}{\alpha_3} - {\varepsilon _3}{\alpha _2}){e^{-2{\alpha_2}a}}}{({\varepsilon_1}{\alpha_2} + {\varepsilon_2}{\alpha_1})({\varepsilon_2}{\alpha_3} + {\varepsilon _3}{\alpha_2})} = 0}
\label{Eq3}
\end{equation}
where $\alpha_{\rm i} = (q^{2}-\epsilon_{\rm i}\omega^{2}/c^{2})^{1/2}$, $q$ is the SPP wavevector, $\epsilon_{1} = \epsilon_{3} = \epsilon_{\rm AlSb} = 11.08$, $\epsilon_{2} = \epsilon_{\rm InSb}$ and $a$ is the thickness of InSb layer. When electrons interact with a high-momentum optical field, the permittivity changes with $\omega$  and $\alpha_{2}$. The permittivity of InSb should therefore be given by a nonlocal Drude model\cite{De2018,SI},
\begin{equation}
{\varepsilon_{\rm nonlocal}}(\omega, q) = {\varepsilon_{\rm \infty}} - \frac{{{\omega _{\rm p}}^2}}{{{\omega ^2} + i\omega \gamma  - (\frac{3}{5} - i\frac{4}{{15}}\frac{\gamma }{\omega })v_F^2{\alpha _2}^2}}	
\label{Eq4}
\end{equation}
where $\epsilon_{\rm \infty}$ is the high-frequency permittivity, $\omega_{\rm p}$ is the plasma frequency, $\gamma$ is the Drude scattering rate, and $v_{\rm F}$ is the Fermi velocity. The energy dispersion of electrons in the conduction band of InSb can be approximately given as $E = (\hbar k)^{2}/2m_{\rm c}$, and $m_{\rm c}$ is the effective electron mass. For a given carrier density $N_{\rm e}$, we obtain $\omega_{\rm p} = \sqrt{N_{\rm e}e^{2}/\epsilon_{0}m_{\rm c}}$, $\gamma = e/(m_{\rm c}\mu_{\rm e})$, $v_{\rm F} = \hbar (3\pi N_{e})^{1/3}/m_{\rm c}$, where $e$ is the electron charge, $\mu_{\rm e}$ is the electron mobility. Here we assume $\mu_{e}$ to be 6000 cm$^{2}$ V$^{-1}$ s$^{-1}$\cite{Litwin1981}. Fig. \ref{Fig. 1}(c) plots the local permittivity as well as the nonlocal
permittivity with $q = 1.2 \times 10^{6}$ cm$^{-1}$. Compared to the local permittivity, the zero-crossing angular frequency $\omega_{\rm ENZ}$ of the nonlocal permittivity is blue-shifted. 
\begin{figure}[!t]
	\begin{center}
	\includegraphics[width=7.5cm]{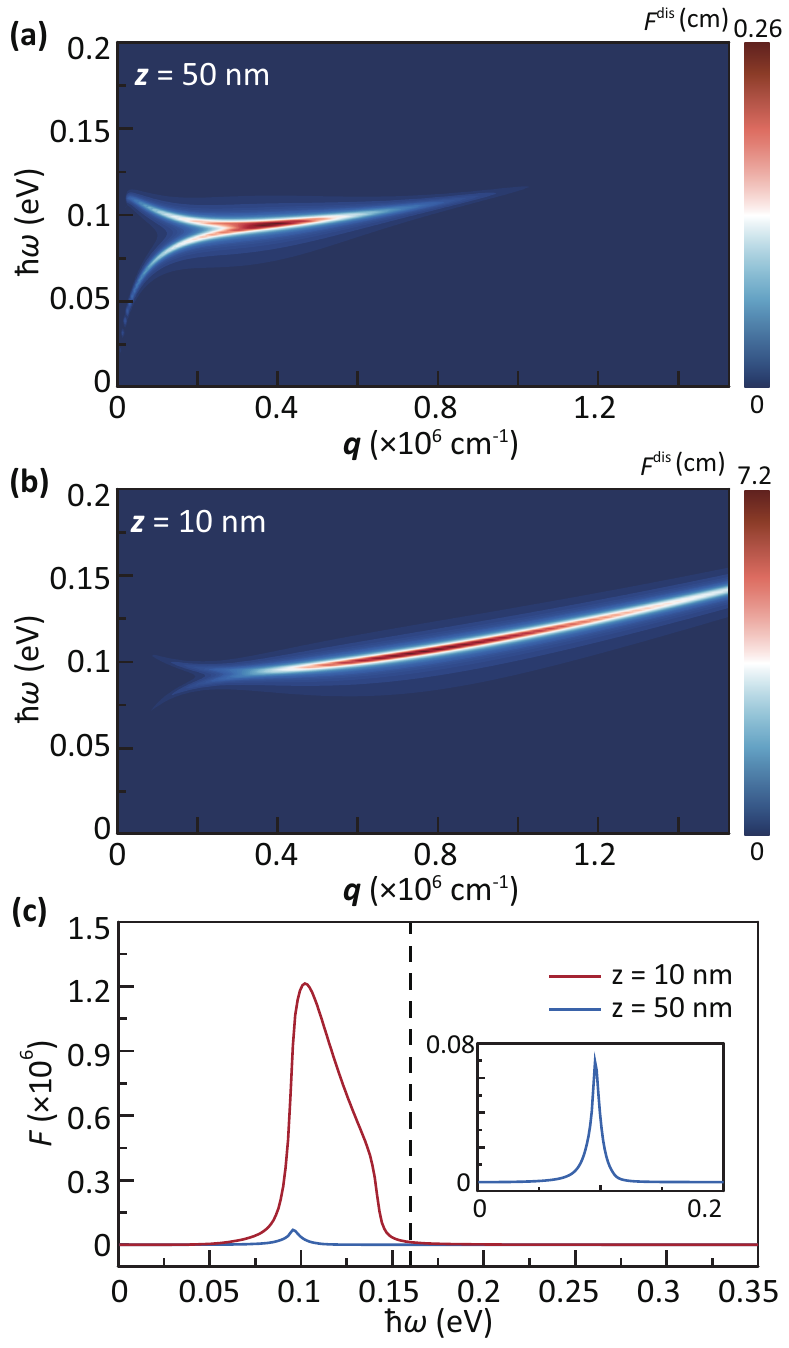}
	\end{center}
	\caption{\label{Fig. 2} (a-b) Distributed Purcell factor $F^{\rm dis}$ as a function of the wavevector and the emitted photon energy at $z$ = 50 nm (a) and 10 nm (b), respectively. (c) Purcell factor $F$ as a function of the emitted photon energy at $z$ = 10 nm and 50 nm. The inset is the zoom-in view of $F$ at $z$ = 50 nm. The black dashed line marks the direct bandgap energy of InSb.}
\end{figure}

To show the influence of the nonlocal effect, we calculate and compare the SPP dispersions with the local and nonlocal permittivity assumption, respectively, as shown in Fig. \ref{Fig. 1}(d). For lossy materials, there is no solutions for	Eq. (\ref{Eq3}) when picking real $q$ and real $\omega$  concurrently. Using complex $q$ or $\omega$ , Eq. (\ref{Eq3}) can find two possible solutions often referred as the long-range (LR)-SPP mode and the short-range (SR)-SPP mode\cite{Wendler1986}. The choice of a real $q$ and consequently a complex $\omega_{q}$ allows the introduction of discrete modes during the mode quantization in our calculation\cite{Archambault2009}, where $\omega_{q}$ represents the complex modal frequency. The local dispersion has a flat asymptote slightly below $\omega_{\rm ENZ}$ and extends to infinite $q$. As the DOS is approximately proportional to $dq/d{\rm Re} (\hbar\omega_{q})$\cite{Archambault2009}, the local permittivity assumption may lead to a diverging Purcell factor near $\omega_{\rm ENZ}$. In contrast, the nonlocal dispersion has a gentle slope near $\omega_{\rm ENZ}$ with a Q-factor about 56, where $Q = {\frac{{\rm Re}(\omega_{ q})}{2{\rm Im}(\omega_{q})}}$. Consequently, a diverging Purcell factor is prevented in our calculation.

Calculating from the dispersion relation (see SI\cite{SI}), the Purcell factor $F$ is given as a funtion of $\omega$ and $z$,
\begin{equation}
{F(\omega ,z) = \int_0^\infty {F^{\rm dis}}(q,\omega ,z){\rm{ }}dq}
\label{Eq5}
\end{equation}
where $F^{\rm dis}$ is the distributed Purcell factor. $F^{\rm dis}$ at $z$ = 10 nm and 50 nm are calculated and shown in Figs. \ref{Fig. 2}(a-c). The spectral broadening of $F^{\rm dis}$ is caused by the finite $Q$-factor of the SPP modes \cite{Iwase2010}. When the emitter is moved from the central to the edge plane of the InSb film, the extremum value of the field confinement near the emitters appears at a higher $q$, leading to a spectral shift of $F$ at different $z$ as illustrated in Fig. \ref{Fig. 2}(d). The maximum Purcell factor can reach $1.2\times10^{6}$ at $z$ = 10 nm, and drops to $7\times10^{4}$ in the center ($z$ = 50 nm).

\begin{figure}[!b]
	\begin{center}
		\includegraphics[width=7.5cm]{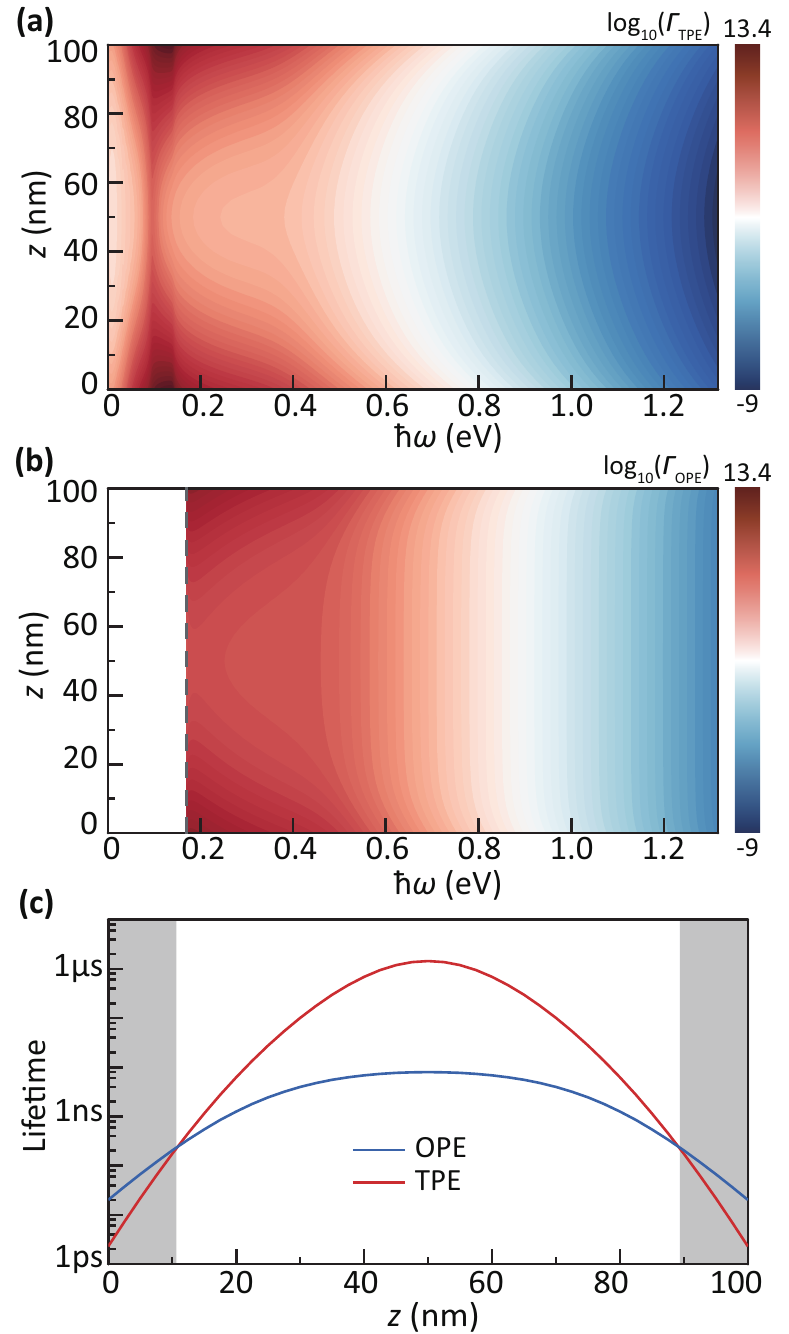}
	\end{center}
	\caption{\label{Fig. 3} (a-b) TPE (a) and OPE (b) rate normalized to $N_{\rm ph}$ in the InSb thin film as a function of the emitted photon frequency and the spatial location. Unit: eV$^{-1}$ $\rm s^{-1}$  (c) Comparison of the TPE and OPE lifetimes as a function of the spatial location in the InSb film. The grey shadowed region marks the spatial location where the TPE lifetime is shorter than the OPE lifetime.}
\end{figure}
\noindent

\section{Result}	
Combining Eqs. (\ref{Eq1}), (\ref{Eq2}) and (\ref{Eq5}), we can calculate the OPE and TPE rate in the InSb film. In the initial calculation, we choose $N_{\rm e} = 8\times10^{18}$ cm$^{-3}$ for InSb, which corresponds to $\omega_{\rm p} = 7\times10^{14}$  rad s$^{-1}$, $m_{\rm c} = 0.05$ $m_{\rm e}$. The intrinsic carrier density in InSb can be neglected in a degenerate doping level. For photoluminescence, we assume a photoexcited carrier density $N_{\rm ph}$ much lower than the doping density, thus avoiding nonlinearity. The TPE and OPE rate in the InSb film are plotted in Fig. \ref{Fig. 3}(a) and \ref{Fig. 3}(b), respectively. While the TPE spectrum starts from the near-zero photon energy, the OPE process only emits photons with energy above the material bandgap. Most importantly, due to the selective Purcell enhancement at frequencies below the material bandgap, TPE can be significantly accelerated in the InSb film. At locations close to the InSb surface, the TPE rate can even surpass the OPE rate. We further retrieve the TPE and OPE lifetime in	the InSb film. TPE can dominate in the range close to the InSb/AlSb interface, as shown in Fig. \ref{Fig. 3}c. Compared with a undoped, bulk InSb material, the 
TPE lifetime in the InSb film can be accelerated from 31 ms to 2.3 ps, with a corresponding TPE/OPE ratio increasing from $4.3\times10^{-7}$ to 10.7.

\begin{figure}[!t]
	\begin{center}
	\includegraphics[width=7.5cm]{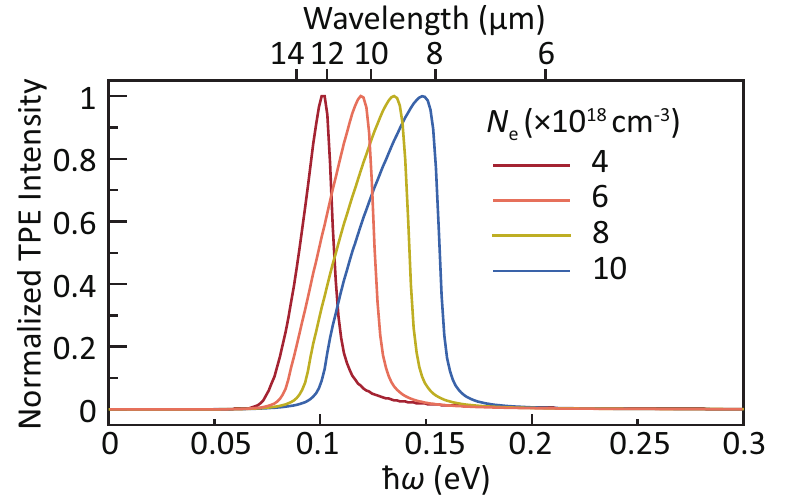}
	\end{center}
	\caption{\label{Fig. 4} Normalized TPE spectrum near the InSb/AlSb interface as a function of $N_{\rm e}$.}
\end{figure}

Moreover, while the spectrum of OPE in a semiconductor, without quantum confinement, is generally fixed by the material bandgap, the spectrum of TPE in a semiconductor can be engineered in a broad range. We further investigate the dependence of the TPE spectrum on the carrier density in the InSb film. As shown in Fig. \ref{Fig. 4}, with an increasing $N_{\rm e}$ from $4\times10^{18}$ cm$^{-3}$ to $1\times10^{19}$ cm$^{-3}$, the spectral maximum of TPE can blue-shift by 45\%, as a result of the modification of the SPP dispersion in the 
InSb film. The tuning of $N_{\rm e}$ can be done by the chemical doping or by applying a static electric 
field. Alternatively, the SPP dispersion can be actively tuned by including additional tunable materials, such as phase-changing vanadium dioxide in the vicinity of the InSb layer\cite{Folland2018}.

\section{Conclusion}	
In conclusion, we theoretically show that ultrafast and tunable TPE can be realized in a subwavelength InSb film, by spatially and spectrally matching TPE with highly-confined SPP modes. Similar concept can be extended to other semiconductors\cite{Charnukha2019}, 2D materials\cite{Rana2011,Kaminer2016}, and superconductors supporting Josephson plasmons\cite{Rajasekaran2016}. The efficiency of TPE can be further boosted by stimulation\cite{Hayat2011}. Ultrafast and efficient TPE holds the potential to enable single and entangled sources emitting at a GHz rate. The polariton emission can be used for realizing plasmonic gain and increasing plasmon coherence, two crucial elements in plasmonic circuits\cite{Basov2017,Heeres2013} and plasmonic lasers\cite{Oulton2009, Fedyanin2012}. The generated polaritons can be coupled into radiative modes by applying near-field momentum compensation through nanotips and nanoantennas\cite{Nevet2010-NL, Charnukha2019}. In addition, the carrier can potentially be injected by electrical means, with possibilities for further on-chip integration.

\section{The Acknowledgements}
This work is supported by the National Natural Science Foundation of China (Grant No. 61975251).

\bibliographystyle{unsrt}
\bibliography{refs}

\newpage
\onecolumn
\begin{center}
	\textbf{\large SUPPLEMENTAL MATERIAL FOR:
		\\Ultrafast two-photon emission in a doped semiconductor thin film}
\end{center}
\setcounter{equation}{0}
\setcounter{figure}{0}
\setcounter{table}{0}
\setcounter{page}{1}
\makeatletter
\renewcommand{\theequation}{S\arabic{equation}}
\renewcommand{\thefigure}{S\arabic{figure}}

\section*{S1. Model of the electronic band structure}
\begin{figure}[h]
	\centering
	\includegraphics[scale=1]{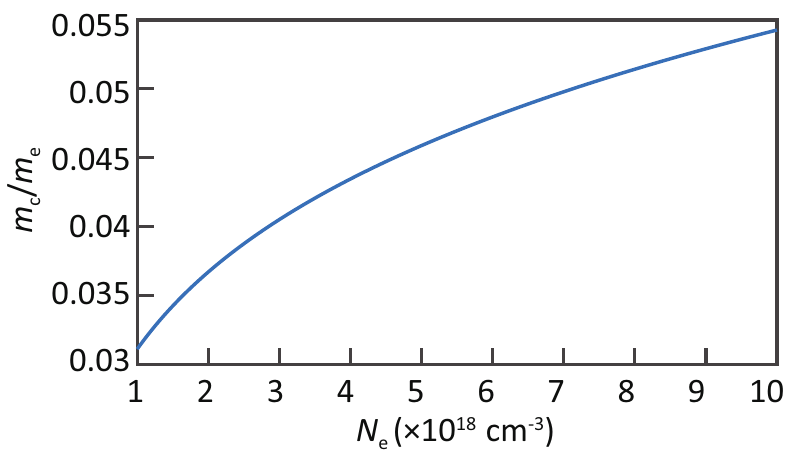}
	\caption{\label{Fig. S1} The effective electron mass $m_{\rm c}$ as a function of the carrier density $N_{\rm e}$ in InSb.}
\end{figure} 
OPE and TPE rates are dependent on the distribution of photo-excited carriers in the conduction and valence bands of InSb. The conduction band structure is decribed as 
$E = {\rm{ }}\frac{{{\hbar ^2}{k^2}}}{{2{m_{\rm c}}}}$. Here, $E$ is the electron energy with respect to the conduction band minimum, $\hbar$ is the reduced Planck's constant, $m_{\rm c}$ is the effective electron mass, and $k$ is the electron wave vector. The value of $m_{\rm c}$ changes with the carrier density $N_{\rm e}$, following the Kane model\cite{Guo2016,Yang2017}. We present $N_{\rm e}$-$m_{\rm c}$ relation in Fig. S1. Photo-excitation also generates holes in the valence band. Most photo-excited holes gather at the heavy-hole valence band\cite{Suhara2004}. The dominant transition is therefore the electron-heavy hole transition. The energy dispersion of the heavy-hole valence band is given by	$E =  - {E_{\rm g}} - \frac{{{\hbar ^2}{k^2}}}{{2{m_{\rm v}}}}$ , where $m_{\rm v}$ = 0.43 $m_{\rm e}$. Here, $m_{\rm e}$ is the electron rest mass. For a given photo-excited carrier density $N_{\rm ph}$ and electronic band structure, the distribution of electrons and holes are decribed by the Fermi-Dirac population functions $f_{\rm c}$ and $f_{\rm v}$, repectively.

\section*{S2. Model of the material permittivity}
The linear optical properties of semiconductors are mainly governed by interband and intraband transitions, and lattice effects\cite{Amirtharaj1994}. The Drude model we use mainly considers intraband contributions. The interband absorption within the OPE and TPE spectra is prohibited by Pauli blocking due to the degenerate doping level in InSb. Furthermore, the optical phonon frequency of InSb\cite{Lockwood2005} is relatively far away from the spectral peak of the modified TPE, thus optical phonon absorption is ignored here. The abovementioned optical properties mainly decribe the local response. When electrons interact with a high-momentum optical field, the dielectric response will change with the momentum of photon, known as the nonlocal effect\cite{Khurgin2015,De2018}. Here, the nonlocal permittivity of InSb is decribed by the nonlocal Drude model.

\section*{S3. Calculation of the Purcell factor in the presence of SPP modes}
Purcell factor describes the enhancement of the spontaneous emission rate. In this work, it is defined as $F(\omega) = \varGamma(\omega)/{\varGamma^0}(\omega )$ , where $\varGamma(\omega)$ and ${\varGamma^0}(\omega )$ are the modified and vacuum emission rate, respectively. The vacuum emission rate is\cite{Khosravi1991,Iwase2010},
\begin{equation}
{\varGamma^0}(\omega ) = {\rm{ }}\frac{{|p|^2{\omega ^3}}}{{3\pi \hbar {\varepsilon _0}{{\rm c}^3}}}
\end{equation}
where $|p|$ is the momentum matrix element of the electric dipole involved in the transition, $\epsilon_0$ is the vacuum permittivity and c is the velocity of light in vacuum. 

In our work, the Purcell enhancement originates from an increased field confinement and squeezed density of states (DOS)\cite{Iwase2010,Shahbazyan2018,Poddubny2012}. Surface plasmon Fourier optics or the Green’s tensor approach can be applied to study Purcell effect\cite{Archambault2009}. Here we adopt the former approach because we can analysis the nonlocal effect directly through the dispersion relation. The electric-field distribution $E(z)$ is an even function of $z$ due to the symmetry of our proposed structure. In analog of the effective mode volume in the resonant cavity, we can use the mode length $L_{\rm q}(z)$ to describe the confinement of the electric field at a given $\omega$ and $q$\cite{Iwase2010,Chen2013},
\begin{equation}
{L_{\rm q}}(z) = \frac{{\int_{-\infty}^{\infty} {d{z_0}{\mathop{\rm Re}\nolimits} [\partial (\omega \varepsilon )/\partial \omega ]|E({z_0}){|^2}} }}{{{\mathop{\rm Re}\nolimits} [\partial (\omega \varepsilon )/\partial \omega ]|E(z){|^2}}}
\end{equation}
where $z_{0}$ is the integral element in $z$-direction. Similar to guided modes in plane waveguides, the electric field of LR-SPP and SR-SPP modes decays in exponential form outside the InSb film, described by Maxwell's equations\cite{Maier2007}. 

The field quantization and mode distribution of SPP modes are fully described by the dispersion relation $q$-$\omega_{ q}$, where $\omega_{q}$ is the complex modal frequency. DOS represents the quantity of SPP modes per energy interval at a given $\omega$. Because the dispersion relation of SPP modes is independent of $z$, DOS is identical for all emitters in the InSb film. The DOS of SPP modes is given by\cite{Iwase2010,Chen2013},
\begin{equation}
D(\omega, q ) = {\frac{1}{\pi }\frac{{{\mathop{\rm Re}\nolimits} ({\omega _{\rm q}})/2Q}}{{{[{\mathop{\rm Re}\nolimits} ({\omega_{q}}) - \omega ]}^2} + [{{{\mathop{\rm Re}\nolimits} ({\omega_{ q}})/2Q]}^2}}}
\end{equation}
where $Q = {\frac{{\rm Re}(\omega_{ q})}{2{\rm Im}(\omega_{q})}}$. The spectral broadening of DOS caused by the finite Q-factor is therefore taken into consideration in Eq. (S3). The average coupling coefficient of the transition dipole moment and SPP modes is $\frac{1}{3}$ by integrating over whole $\textbf{\textsl{q}}$ space. The distributed Purcell factor $F^{\rm dis}$ is therefore expressed as\cite{Iwase2010},
\begin{equation}
{F^{\rm dis}}(\omega, q, z){\rm{ }} = {\rm{ }}\frac{{\pi {\rm{ }}{c^3}\kappa qD(\omega )}}{{{L_{\rm q}}(z){\omega ^2}}}
\end{equation}
where $\kappa$ is the ratio of the electric field energy to the total field energy. The Purcell factor $F$ is the integral of $F^{\rm dis}$ over $q$, 
\begin{equation}
F\left( {\omega ,z} \right){\rm{ }} = \int_0^\infty {F^{\rm dis}}\left( {\omega ,q,z} \right){\rm{ }}dq
\end{equation}

\section*{S4. Calculation of the OPE and TPE rates}
In this section, we give a brief derivation on the modified OPE and TPE rates in semiconductors. The vaccum OPE and TPE processes are modified by the electric polarization in undoped InSb or SPP modes in degenerately doped InSb. Here, ${\varGamma_{\rm OPE}^0}(\omega )$, in the unit of s$^{-1}$, is used to represent the emission rate for one given initial electron state coupled to vacuum electromagnetic states. To better compare these two dielectric environments, we start from the vacuum emission rate in a given semiconductor volume $V$. The unmodified OPE rate per unit volume $R_{\rm OPE}^0$ at a given ${\omega _0}$, in the unit of eV${^{-1}}$ cm$^{-3}$ s$^{-1}$, can be obtained by summing all possible initial states in the first Brillouin zone (FBZ)\cite{Basu1997}, 
\begin{equation}
\begin{aligned}
R_{\rm OPE}^0({\omega _0}) &= \frac{2}{V}\sum\limits_{k} {{\varGamma _{\rm OPE}^0}({\omega _0})}{\delta _{{k_{\rm c}},k}}\delta [{E_{\rm c}}({k}) - {E_{\rm v}}({k}) - \hbar \omega_0 ] {f_c}{f_v}
\\&= 2 \times \int\limits_{\rm FBZ} {\frac{{{d^3}k}}{{{{(2\pi )}^3}}}} {\varGamma_{\rm OPE} ^0}({\omega _0}){\delta _{{k_{\rm c}},k}}\delta [{E_{\rm c}}({k}) - {E_{\rm v}}({k}) - \hbar \omega_0 ]{f_{\rm c}}{f_{\rm v}}
\\&= \frac{{ < |{p_{\rm cv}}{|^{^2}} > \omega _0^3}}{{3\pi \hbar {\varepsilon _0}{c^3}}}J(\hbar {\omega _0}){f_{\rm c}}{f_{\rm v}}
\end{aligned}
\end{equation}
The average value of the squared momentum matrix element $<|{p_{\rm cv}}{|^{^2}}>$ is equal to $\frac{m_{\rm e}E_{\rm p}}{6}$, where the Kane's energy $E_{\rm p}$ of InSb is 23.3eV\cite{Baimuratov2013}. The joint density of state $J(\hbar {\omega})$ is defined as\cite{Basu1997},
\begin{equation}
J(\hbar {\omega}) = 2 \times \int\limits_{\rm FBZ} {\frac{{{d^3}k}}{{{{(2\pi )}^3}}}} \delta ({E_{\rm c}}({k}) - {E_{\rm v}}({k}) - \hbar {\omega}) = \frac{1}{{2{\pi ^2}}}{(\frac{{2{m_{\rm r}}}}{{{\hbar ^2}}})^{3/2}}\sqrt {\hbar {\omega} - {E_{\rm g}}} 
\end{equation}
where the reduced carrier mass $m_{\rm r}$ is determined by $\frac{1}{{{m_{\rm r}}}} = \frac{1}{{{m_{\rm c}}}} + \frac{1}{{{m_{\rm v}}}}$. $J{f_{\rm c}}{f_{\rm v}}$ gives the number of available electron-hole pairs per unit volume per unit energy interval. In the InSb film, if emitted photons are coupled to SPP modes with different $q$, the modified emission rate $\varGamma$ becomes a function of $\omega$ and $q$. The modified OPE rate per unit volume at a given ${\omega _0}$ can be expressed as,
\begin{equation}
\begin{aligned}
{R_{\rm OPE}}(\omega_0 ) &= \frac{2}{V}\sum\limits_{{k}} {\varGamma_{\rm OPE} (\omega_0 ,q)}{\delta _{{k_{\rm c}},k}}\delta [{E_{\rm c}}({k}) - {E_{\rm v}}({k}) - \hbar \omega_0 ] {f_{\rm c}}{f_{\rm v}} \\ 
&= 2 \times \int\limits_{\rm FBZ} {\frac{{{d^3}{k}}}{{{{(2\pi )}^3}}}} \int_0^\infty {dq{F^{\rm dis}}(\omega_0 ,q)} \frac{{|{p_{\rm cv}}{|^{^2}}{\omega_0 ^3}}}{{3\pi \hbar {\varepsilon _0}{c^3}}}{\delta _{{k_{\rm c}},k}}\delta [{E_{\rm c}}({k}) - {E_{\rm v}}({k}) - \hbar \omega_0 ]{f_{\rm c}}{f_{\rm v}} \\ 
&= \int_0^\infty {dq{F^{\rm dis}}(\omega_0 ,q)} \frac{{ < |{p_{\rm cv}}{|^{^2}} > {\omega_0 ^3}}}{{3\pi \hbar {\varepsilon _0}{c^3}}}J(\hbar \omega_0 ){f_{\rm c}}{f_{\rm v}} \\ 
&= F(\omega_0 )\frac{{ < |{p_{\rm cv}}{|^{^2}} > {\omega_0 ^3}}}{{3\pi \hbar {\varepsilon _0}{c^3}}}J(\hbar \omega_0 ){f_{\rm c}}{f_{\rm v}} \\ 
&= F(\omega_0 )R_{\rm OPE}^0
\end{aligned}
\end{equation}	
The spectrally-integrated OPE rate ${W_{\rm OPE}}$, in the unit of cm$^{-3}$ s$^{-1}$, is 
\begin{equation}
{W_{\rm OPE}} = \int_0^\infty {R_{\rm OPE}}d\hbar \omega
\end{equation}
The OPE lifetime in average is defined as
\begin{equation}
{\tau _{\rm OPE}} = \frac{{{N_{\rm ph}}}}{{{W_{\rm OPE}}}}
\end{equation}

Spontaneous TPE is a second-order process in the perturbation theory\cite{Basu1997,Andersson1985}. Similar to two-photon absorption, TPE can be theoretically treated as two concurrent first-order transitions: from the initial- to the intermediate-state, and from the intermediate- to the final-state. All Bloch functions or excitonic states have the possibility functioning as the intermediate states. In practice, it is usually necessary to limit the whole intermediate state set to a limited subset, which provides the main contribution to the transition probability.  The initial and final states themselves are usually considered intermediate states\cite{Andersson1985,Hayat2009}. TPE rate in vacuum is therefore proportional to the multiplication of the emission rates of these two concurrent first-order transitions\cite{Andersson1985,Hayat2009},
\begin{equation}
\begin{aligned} 
R_{\rm TPE}^0({\omega _1},{\omega _2}){\text{ }} &= {\text{ }}\frac{\hbar }{{2\pi }}{(\frac{{{m_{\rm c}}}}{{{m_{\rm r}}}})^2}\sum\limits_i {(\frac{{|{p_{\rm ci}}{|^{^2}}{\omega _1}^3}}{{3\pi \hbar {\varepsilon _0}{c^3}}}\frac{{|{p_{\rm iv}}{|^{^2}}{\omega _2}^3}}{{3\pi \hbar {\varepsilon _0}{c^3}}})} M \\ 
&= \frac{\hbar }{{2\pi }}{(\frac{{{m_{\rm e}}}}{{{m_{\rm r}}}})^2}\frac{{(|{p_{\rm cc}}{|^{^2}} + |{p_{\rm vv}}{|^{^2}}){\omega _1}^3}}{{3\pi \hbar {\varepsilon _0}{c^3}}}\frac{{|{p_{\rm cv}}{|^{^2}}{\omega _2}^3}}{{3\pi \hbar {\varepsilon _0}{c^3}}}M \\
\end{aligned} 
\end{equation}
Specifically, $\frac{{|{p_{\rm ci}}{|^{^2}}{\omega _1}^3}}{3\pi \hbar {\varepsilon _0}{c^3}}$ can be understood as the occurrence rate of the initial-intermediate transition, while  $\frac{{|{p_{iv}}{|^{^2}}{\omega _2}^3}}{{3\pi \hbar {\varepsilon _0}{c^3}}}$ can be understood as the occurrence rate of the intermediate-final transition, where $\omega_1$ and $\omega_2$ are exchangeable. Two photons $\omega_1$, $\omega_2$ are emitted simultaneously during TPE. The initial-intermediate transition may emit at a angular frequency of either $\omega_1$ or $\omega_2$, leading to the coefficient $M = {\left| {\frac{1}{{\hbar {\omega _1} + i\hbar {\gamma _{\rm d}}}} + {\text{ }}\frac{1}{{\hbar {\omega _2} + i\hbar {\gamma _{\rm d}}}}} \right|^2}$, where $\gamma_{\rm d}$ is the dephasing rate\cite{Hayat2009}. $|p_{\rm cc}|$ and $|p_{\rm vv}|$ are the momentum eigenvalues of the initial and final states, respectively. Eq. (S11) is further expressed as,
\begin{equation}
\varGamma_{\rm TPE}^0({\omega _1},{\omega _2}){\text{ }} = {\text{ }}\frac{\hbar }{{\pi }}{(\frac{{{m_{\rm e}}}}{{{m_{\rm r}}}})^2}\frac{{{{(\hbar {k_{\rm c}})}^2}{\omega _1}^3}}{{3\pi \hbar {\varepsilon _0}{c^3}}}\frac{{|{p_{\rm cv}}{|^{^2}}{\omega _2}^3}}{{3\pi \hbar {\varepsilon _0}{c^3}}}M
\end{equation}
where $k_{\rm c}$ is the electron wave vector of the initial state. Similar to the derivation of the OPE rate, the vacuum TPE rate per unit volume of the semiconductor can be obtained by summing all possible initial states in the first Brillouin zone (FBZ), 
\begin{equation}
\begin{aligned}
R_{\rm{TPE}}^0({\omega _1},{\omega _2}) &= \frac{2}{V}\sum\limits_{{k}} {\varGamma _{\rm{TPE} }^0({\omega _1},{\omega _2}}){\delta _{{k_{\rm c}},k}}\delta [{E_{\rm c}}(k) - {E_{\rm v}}(k) - \hbar {\omega _1} - \hbar {\omega _2}]{f_{\rm c}}{f_{\rm v}} \\ 
&= 2 \times \int\limits_{\rm{FBZ} } {\frac{{{d^3}k}}{{{{(2\pi )}^3}}}} \frac{\hbar }{{\pi }}{(\frac{{{m_{\rm e}}}}{{{m_{\rm r}}}})^2}\frac{{{{(\hbar k)}^2}{\omega _1}^3}}{{3\pi \hbar {\varepsilon _0}{c^3}}}\frac{{|{p_{\rm cv}}{|^{^2}}{\omega _2}^3}}{{3\pi \hbar {\varepsilon _0}{c^3}}}M{\delta _{{k_{\rm c}},k}}\delta [{E_{\rm c}}(k) - {E_{\rm v}}(k) - \hbar {\omega _1} - \hbar {\omega _2}]{f_{\rm c}}{f_{\rm v} } \\ 
&= \varGamma _{\rm{TPE} }^0({\omega _1},{\omega _2})MJ(\hbar {\omega _1} + \hbar {\omega _2}){f_{\rm c}}{f_{\rm v}} \\ 
\end{aligned}
\end{equation}
The simultaneously-emitted two photons are coupled to SPP modes in the InSb film. The modified emission rate $\varGamma _{\rm{TPE} }$ therefore becomes a function of ${\omega _1}$, ${q_1}$, ${\omega _2}$ and ${q_2}$,
\begin{equation}
\begin{aligned}
{R_{\rm{TPE} }}({\omega _1},{\omega _2}) &= \frac{2}{V}\sum\limits_{{k}} {{\varGamma _{\rm{TPE} }}({\omega _1},{q_1},{\omega _2},{q_2})}{\delta _{{k_{\rm c}},k}}\delta [{E_{\rm c}}(k) - {E_{\rm v}}(k) - \hbar {\omega _1} - \hbar {\omega _2}] {f_{\rm c}}{f_{\rm v}} \\ 
&= 2 \times \int\limits_{\rm{FBZ} } {\frac{{{d^3}k}}{{{{(2\pi )}^3}}}}{\int_0^\infty {d{q_1}{F^{\rm{dis} }}({\omega _1},{q_1})\int_0^\infty {dq_{2}} }}{F^{\rm dis}}({\omega_2},{q_2})\frac{\hbar }{{\pi }}{(\frac{{{m_{\rm e}}}}{{{m_{\rm r}}}})^2}\frac{{{{(\hbar k)}^2}{\omega _1}^3}}{{3\pi \hbar {\varepsilon _0}{c^3}}}\frac{{|{p_{cv}}{|^{^2}}{\omega _2}^3}}{{3\pi \hbar {\varepsilon _0}{c^3}}}M\\
&\quad \ {\delta _{{k_{\rm c}},k}} \delta [{E_c}(k) - {E_v}(k) - \hbar {\omega _1} -\hbar {\omega _2}]{f_{\rm c}}{f_{\rm v}} \\ 
&= F({\omega _1})F({\omega _2})\varGamma _{\rm{TPE} }^0({\omega _1},{\omega _2})MJ(\hbar {\omega _1} + \hbar {\omega _2}){f_{\rm c}}{f_{\rm v}} \\ 
&= F({\omega _1})F({\omega _2})R_{\rm{TPE} }^0({\omega _1},{\omega _2}) 
\end{aligned}
\end{equation}
The spectrally-integrated TPE rate is 
\begin{equation}
W_{\rm TPE} = {\text{ }}{\frac{1}{2}}\int_0^\infty\int_0^\infty {R_{\rm TPE}}d\hbar {\omega _1}d\hbar {\omega _2}
\end{equation}
where 1/2 is from the exchangeability of $\omega_1$ and $\omega_2$. Note that ${R_{\rm TPE}}$ is equal to zero when $(\hbar {\omega _1}+\hbar {\omega _2})$ is below $E_{\rm g}$.  The TPE lifetime in average can be given by,
\begin{equation}
{\tau_{\rm TPE}} = {N_{\rm ph}}/(\frac{1}{2}{W_{\rm TPE}})
\end{equation}
where 1/2 is due to the simultaneous emission of two photons during TPE.


\end{document}